\documentclass[preprint,review,english,american,aps,reprint,superscriptaddress,12pt]{revtex4}
\usepackage[LGR,T1]{fontenc}
\usepackage[latin9]{inputenc}
\usepackage{amsmath}
\usepackage{amssymb}
\usepackage{graphicx}
\usepackage{textcomp}
\setcitestyle{super}
\usepackage{physics}

\makeatletter

\ProvideTextCommand{\~}{LGR}[1]{\char126#1}

\@ifundefined{textcolor}{}
{%
 \definecolor{BLACK}{gray}{0}
 \definecolor{WHITE}{gray}{1}
 \definecolor{RED}{rgb}{1,0,0}
 \definecolor{GREEN}{rgb}{0,1,0}
 \definecolor{BLUE}{rgb}{0,0,1}
 \definecolor{CYAN}{cmyk}{1,0,0,0}
 \definecolor{MAGENTA}{cmyk}{0,1,0,0}
 \definecolor{YELLOW}{cmyk}{0,0,1,0}
}

\ProvideTextCommand{\~}{LGR}[1]{\char126#1}

\@ifundefined{textcolor}{}{%
 \definecolor{BLACK}{gray}{0}
 \definecolor{WHITE}{gray}{1}
 \definecolor{RED}{rgb}{1,0,0}
 \definecolor{GREEN}{rgb}{0,1,0}
 \definecolor{BLUE}{rgb}{0,0,1}
 \definecolor{CYAN}{cmyk}{1,0,0,0}
 \definecolor{MAGENTA}{cmyk}{0,1,0,0}
 \definecolor{YELLOW}{cmyk}{0,0,1,0}
}

\usepackage[english]{babel}

\def\url#1{}
\addto\captionsenglish{%
}

\usepackage[none]{hyphenat}

\usepackage{babel}

\usepackage{babel}

\usepackage{babel}

\makeatother

\usepackage{babel}
\begin{document}

\title{A van der Waals Interface Hosting Two Groups of Magnetic Skyrmions}

\author{Yingying Wu}
\thanks{Correspond to: yywu@mit.edu}
\affiliation{Francis Bitter Magnet Laboratory and Plasma Science and Fusion Center, Massachusetts Institute of Technology, Cambridge, MA 02139, USA}

\author{$^{,\dagger}$ Brian Francisco}
\thanks{These authors contributed equally to this work }
\affiliation{Department of Physics and Astronomy, University of
California, Riverside, CA 92521, USA}

\author{Wei Wang}

\affiliation{Key Laboratory of Flexible Electronics \& Institute of Advanced Materials, Jiangsu National Synergetic Innovation Center for Advanced Materials, Nanjing Tech University, Nanjing 211816, China}

\author{Yu Zhang}
\author{Caihua Wan}
\author{Xiufen Han}

\affiliation{Institute of Physics, Chinese Academy of Sciences, Beijing, 100190, China}

\author{Hang Chi}
\affiliation{Francis Bitter Magnet Laboratory and Plasma Science and Fusion Center, Massachusetts Institute of Technology, Cambridge, MA 02139, USA}
\affiliation{U.S. Army CCDC Army Research Laboratory, Adelphi, MD 20783, USA }
\author{Yasen Hou}
\author{Alessandro Lodesani}
\affiliation{Francis Bitter Magnet Laboratory and Plasma Science and Fusion Center, Massachusetts Institute of Technology, Cambridge, MA 02139, USA}

\author{Yong-tao Cui}

\affiliation{Department of Physics and Astronomy, University of
California, Riverside, CA 92521, USA}

\author{Kang L. Wang}


\affiliation{Department of Electrical and Computer Engineering, University of
California, Los Angeles, Los Angeles, CA 90095, USA.}

\author{Jagadeesh S. Moodera}
\thanks{Correspond to: moodera@mit.edu }
\affiliation{Francis Bitter Magnet Laboratory and Plasma Science and Fusion Center, Massachusetts Institute of Technology, Cambridge, MA 02139, USA}
\affiliation{Physics Department, Massachusetts Institute of Technology, Cambridge, MA 02139, USA}


\begin{abstract}
Multiple magnetic skyrmion phases add an additional degree of freedom for skyrmion based ultrahigh-density spin memory devices. Extending the field to two-dimensional van der Waals magnets is a rewarding challenge, where the realizable degree of freedoms (e.g. thickness, twisting angle and electrical gating) and high skyrmion density result in intriguing new properties and enhanced functionality. We report a van der Waals interface, formed by two 2D ferromagnets Cr$_2$Ge$_2$Te$_6$ and Fe$_3$GeTe$_2$ with a Curie temperature of $\sim$ 65 K and $\sim$ 205 K, respectively, hosting two groups of magnetic skyrmions.  
Two sets of topological Hall effect are observed below 60 K when Cr$_2$Ge$_2$Te$_6$ is magnetically ordered. These two groups of skyrmions are directly imaged using magnetic force microscopy. 
Interestingly, the magnetic skyrmions persist in the heterostructure in the remanent state with zero applied magnetic field. Our results are promising for the realization of skyrmionic devices based on van der Waals heterostructures hosting multiple skyrmion phases.
\end{abstract}

\maketitle

Two dimensional (2D) magnetism was recently discovered in van der Waals (vdW) ferromagnets\cite{gong2017discovery, deng2018gate,fei2018two} and antiferromagnets\cite{huang2017layer,song2018giant,deng2020quantum}, providing unprecedented opportunities for exploring magnetism, and towards spintronic applications in the 2D limit\cite{burch2018magnetism,avsar2020colloquium}. 
Among all the interface engineered heterostructures based on vdW layered systems, magnetic proximity effect is integral to manipulating spintronic\cite{tang2020magnetic,wu2020large,dankert2017electrical}, superconducting\cite{kim2017strong,wu2019induced,lupke2020proximity} and topological\cite{kezilebieke2020topological,island2019spin,hellman2017interface} phenomena. Topological magnetic skrymions have been well studied due to their nontrivial topology, which lead to many interesting fundamental and dynamical properties\cite{nagaosa2013topological, fert2013skyrmions,fert2017magnetic}. 
These have been reported mostly for non-centrosymmteric single crystals\cite{muhlbauer2009skyrmion,yu2010real,jonietz2010spin}, ultrathin epitaxial system\cite{wang2018ferroelectrically,woo2016observation}, and magnetic multilayers\cite{jiang2015blowing,li2019anatomy,zeissler2020diameter,wiesendanger2016nanoscale,maccariello2018electrical}. 
Recently N\'eel-type skyrmions were observed in a vdW ferromagnet interfaced with an oxidized layer\cite{park2021neel} or a transition metal dichalcogenide\cite{wu2020neel}  with a control of the skyrmion phase through tuning of the ferromagnet thickness. Furthermore, with a variety of vdW magnets, skrymions phase could be created in their new interfaces with unique properties. 

Materials hosting multiple skyrmion phases add richness to the field, with an additional degree of freedom in designing devices. Such a degree of freedom adds value as a means to lower the error rate in racetrack memories with skyrmions as information carriers\cite{mandru2020coexistence,suess2018repulsive}. Multiple skyrmion phases have been reported in FeGe grown on Si(111) at low temperatures\cite{ahmed2018chiral} and in Ir/Fe/Co/Pt multilayers\cite{mandru2020coexistence}. Different from these thin film systems, vdW heterostructures possess well defined and ordered atomic layer range of thicknesses, atomically sharp interfaces and easy electrical field control. Forming a heterostructure of two vdW ferromagnets allows breaking of inversion symmetry at the interface\cite{fert2013skyrmions}, with skyrmions expected inside both these two ferromagnetic layers in the context of strong spin orbit coupling. So far, there are no direct observations of the multi groups of magnetic skyrmions at a vdW interface.          

In this work, we demonstrate a vdW interface, i.e. Fe$_3$GeTe$_2$/Cr$_2$Ge$_2$Te$_6$ (FGT/CGT) interface, can host two groups of magnetic skyrmions. The initial signature for this comes from the observation of the topological Hall effect (THE). Below the Curie temperature $T_\textrm{C}$ ($\sim$ 65 K) of CGT, Hall resistivity shows two sets of kinks,  due to the magnetic skyrmion phase formation on CGT and FGT sides. Above 60 K, there is only one set of kinks seen in the Hall resistivity, likely arising from the skyrmion phase on the FGT side. To ascertain and understand the origin of THE, magnetic force microscopy was used to image  the skyrmion lattice directly. These observations could provide a new platform for designing a robust skyrmion-based memory and computing architectures by coding the information in multiple groups of skyrmions.

\subsection*{Ferromagnetic Fe$_3$GeTe$_2$ and Cr$_2$Ge$_2$Te$_6$ thin layers}
\begin{figure}
\centering{}\includegraphics[width=0.95\columnwidth]{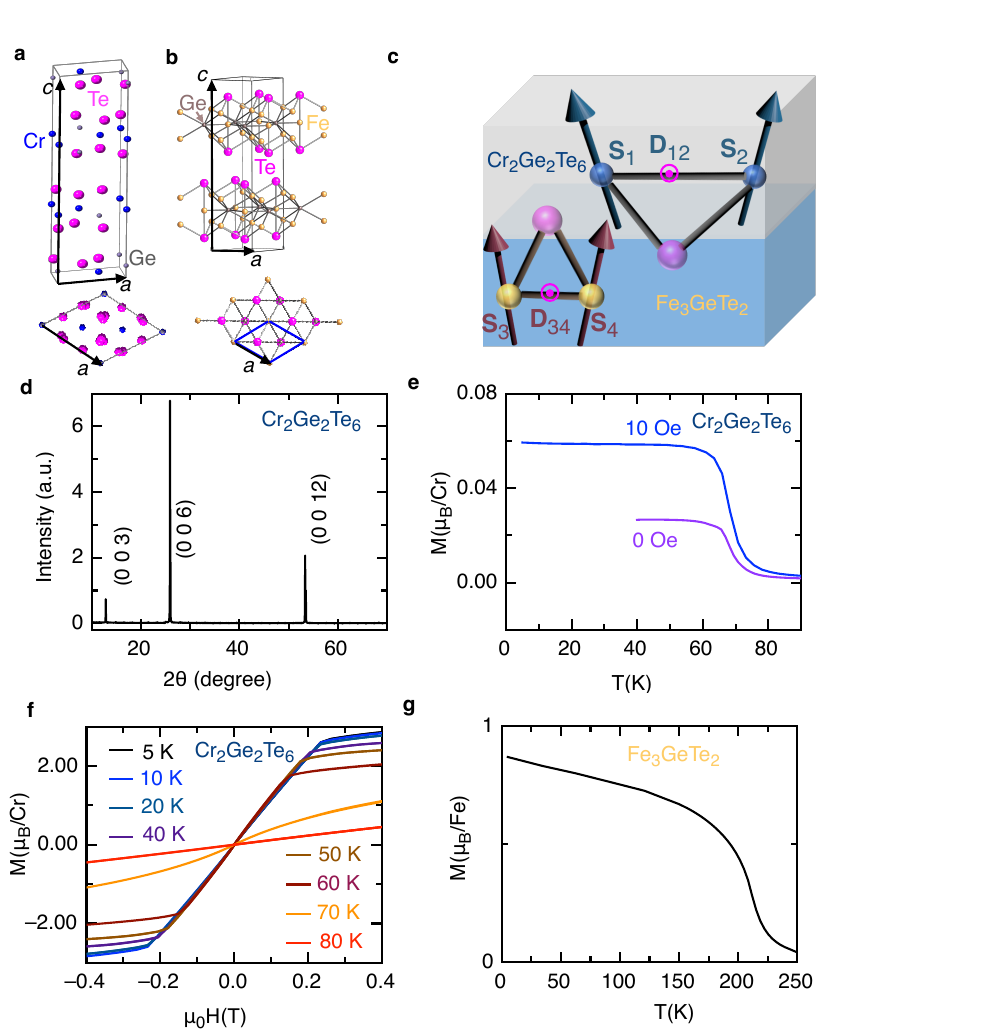}
\caption{\textbf{Structure and magnetic properties in FGT and CGT crystals.} \textbf{a}, Crystal structure of CGT. Side view and top view for the rhombohedral structure.  \textbf{b}, Crystal structure of FGT. \textbf{c}, DMI is induced at the interface of CGT and FGT. \textbf{d}, X-ray diffraction pattern of CGT indicating high quality crystal. \textbf{e}, The temperature dependent magnetization profile of CGT under zero-field-cool and field-cool (out-of-plane applied magnetic field $H$ = 10 Oe) conditions, revealing a $T_\textrm{C}$ of $\sim$ 65 K.  \textbf{f}, Magnetic hysteresis profile of CGT at selected temperatures with applied magnetic field $H\,//\,c$-axis displaying an anisotropy field of approximately 220 mT at 20 K. \textbf{g}, The temperature dependent magnetization profile of FGT under zero-field cooling condition, indicating a $T_\textrm{C}$ of $\sim$ 205 K.  \label{Fig:Mag} }
\end{figure}

CGT is a soft, 2D layered ferromagnetic crystal, with rhombohedral structure and vdW stacking of hexagonal layers. As illustrated in Fig. \ref{Fig:Mag}a, a layer of Cr atoms is sandwiched between two Ge-Te layers, forming a monolayer of CGT. Its lattice parameters are $a$ = 6.826 \AA \,and $c$ = 20.531 \AA. Among the 2D magnets, FGT is a ferromagnet with a strong perpendicular magnetic anisotropy. Bulk FGT consists of weakly bonded Fe$_3$Ge layers that alternate with two Te layers with a space group P6$_3$mmc as shown in Fig. \ref{Fig:Mag}b with lattice parameters $a$ = 3.991 \AA \,and $c$ = 16.333 \AA\cite{yi2016competing}. For both pristine CGT and FGT crystals, the inversion symmetry is preserved inside the bulk. However, by forming an interface between these two materials, the inversion symmetry is broken. With spin-orbit coupling present, Dzyaloshinskii-Moriya interaction (DMI) is induced at the interface as shown in Fig. \ref{Fig:Mag}c, both at FGT and CGT interfacial layers. The DMI between two atomic spins $\vb*{S}_1$ and $\vb*{S}_2$ can be expressed as: $H_{\textrm{DM}}=-\vb*{D}_{12}\cdot \vb*{S}_1\times \vb*{S}_2$.

The crystal structure of grown CGT crystal was analyzed by using X-ray diffraction studies as depicted in Fig. \ref{Fig:Mag}d. Typical peaks corresponding to the (003), (006) and (0012) planes in CGT were observed. These peaks indicate that the layered CGT crystal with good quality is achieved. The $T_\textrm{C}$ for CGT was around 65 K, as seen from magnetization measurement using a magnetometer(Fig. \ref{Fig:Mag}e). The ferromagnetic order in CGT develops at low temperatures upon cooling down both without and with an out-of-plane magnetic field. The magnetic hysteresis is shown in Fig. \ref{Fig:Mag}f, where the ferromagnetism disappears above 65 K. Compared to CGT, FGT has a much higher $T_\textrm{C}$ of around 200 K, as shown in Fig. \ref{Fig:Mag}g.

\subsection*{Two sets of THE signals}
THE can arise from the accumulated Berry phase acquired by electrons in the adiabatic limit as they encounter a skyrmion along their path. However, recent experiments offered an alternative interpretation of the topological Hal effect reported in topological insulator heterostructures; they attribute the topological Hall signals to the overlapping of two anomalous Hall signals with opposite signs. A similar reasoning was used to explain the topological Hall-like signatures (one peak and one dip) in SrTiO$_3$/SrRuO$_3$/SrTiO$_3$\cite{kimbell2020two} heterostructures. In such a scenario, the competing anomalous Hall signals contributing to topological Hall-like signatures originate either from the coexisting surface and bulk magnetic phases in the magnetic topological insulator system \cite{fijalkowski2020coexistence}, or two interfaces in the heterostructures. As opposed to previous systems, this current FGT/CGT bilayer system might facilitate the demonstration of THE for three reasons: (1) only one interface is formed between the two materials; (2) two sets of kinks, a signature for THE at both sides of an interface, are observed when the temperature is lower than the $T_\textrm{C}$ of CGT; and (3) skyrmions are imaged directly by magnetic force microscopy (MFM).  

\begin{figure}
\centering{}
\includegraphics[width=0.85\columnwidth]{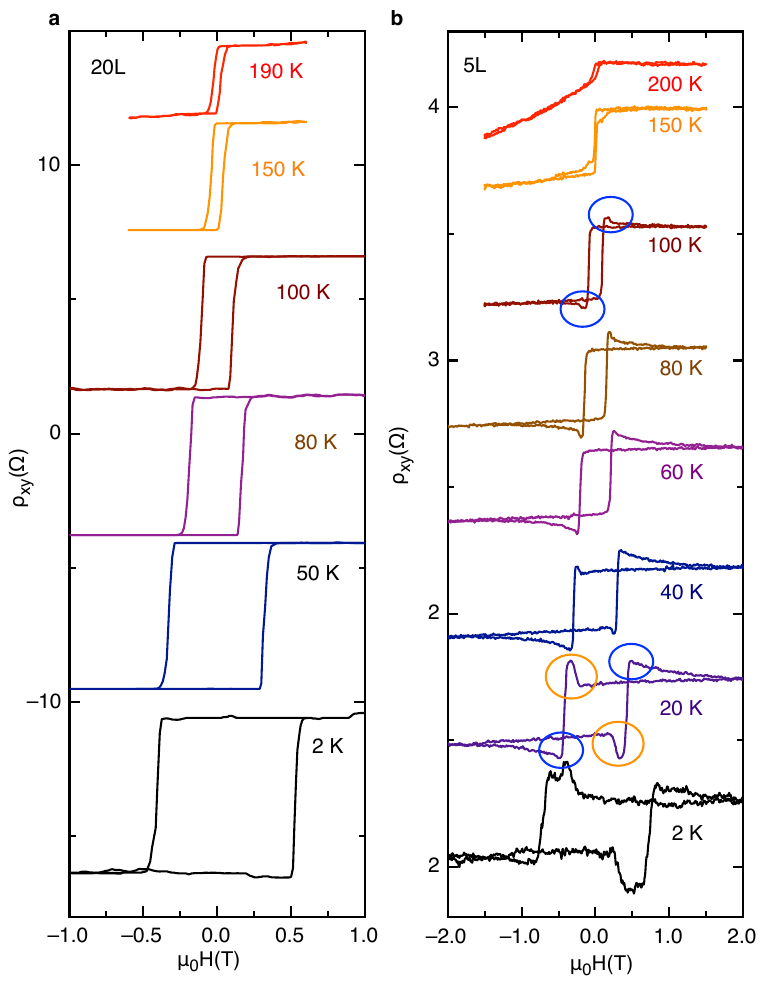}
\caption{\textbf{Topological Hall effect in the Cr$_2$Ge$_2$Te$_6$/Fe$_3$GeTe$_2$ heterostructure}. Hall resistivity (shifted for clarity) of a CGT/FGT heterostructure with different FGT thickness of \textbf{a}, 20L,  and \textbf{b}, 5L. A magnetic field was applied perpendicular to the sample plane, i.e., along the $c$-axis. A square hysteresis loop indicates the perpendicular magnetic anisotropy. Peaks and dips appear in heterostructure with 5L FGT, signifying the existence of topological Hall effect. \label{Fig:THE}}
\end{figure}

In the transport measurements, CGT/FGT thin films were transferred onto prepared Hall-bar bottom electrodes under a microscope (see Supplementary Information Section A for device images) inside the glove box (N$_2$ atmosphere), and the resistivity was measured with lock-in technique. The CGT layers for the transport measurement have a thickness between 30 nm and 40 nm, which is relatively insulating compared to metallic FGT layers. Magnetic field was applied perpendicular to the sample plane. As shown in Fig. \ref{Fig:THE}a, 20-layer (20L)  FGT exhibits square hysteresis loop when the temperature is below 200 K, demonstrating a strong perpendicular out-of-plane magnetic anisotropy. When FGT thickness was reduced to 5L, interface coupling between FGT and CGT gave rise to pronounced peaks and dips near the magnetic transition edge, the signature of THE. Different from previously reported systems, two sets of THEs (as indicated by blue and orange circles in Fig. \ref{Fig:THE}b) are observed in this heterostructure of two ferromagnets. 

The topological Hall resistivity value is related to carrier density $n$, and an effective field $B_{\textrm{eff}}$ generated by skyrmion lattice ($\rho_{\textrm{xy}}^{\textrm{THE}}\sim\frac{1}{ne}B_{\textrm{eff}}$)\cite{neubauer2009topological}. The different magnitude of the topological Hall resistivity value indicates two groups of skyrmions with different sizes. These two sets of THEs coexist when the temperature is below the $T_\textrm{C}$ of CGT. One set indicated by the orange circles vanishes at 60 K and above, is attributable to the disappearance of ferromagnetic order in CGT with increasing temperature. It may be noted that, this THE is only observed in CGT/5L FGT sample. The absence of THE in this heterostructure with thicker FGT can be due to the high carrier density. With larger carrier density in thicker FGT samples, the topological Hall resistivity is dwarfed in comparison. For example, for 20L FGT at 20 K, the estimated topological Hall resistivity contribution is less than $10^{-3} \,\Omega$, assuming a skyrmion lattice with a size of 100 nm. This resistivity value is more than two orders of magnitude smaller than the anomalous Hall resistivity, and thus tends to be buried in the anomalous Hall signals.

\subsection*{Magnetic skyrmions directly imaged by MFM}

\begin{figure}
\centering{}
\includegraphics[width=1\columnwidth]{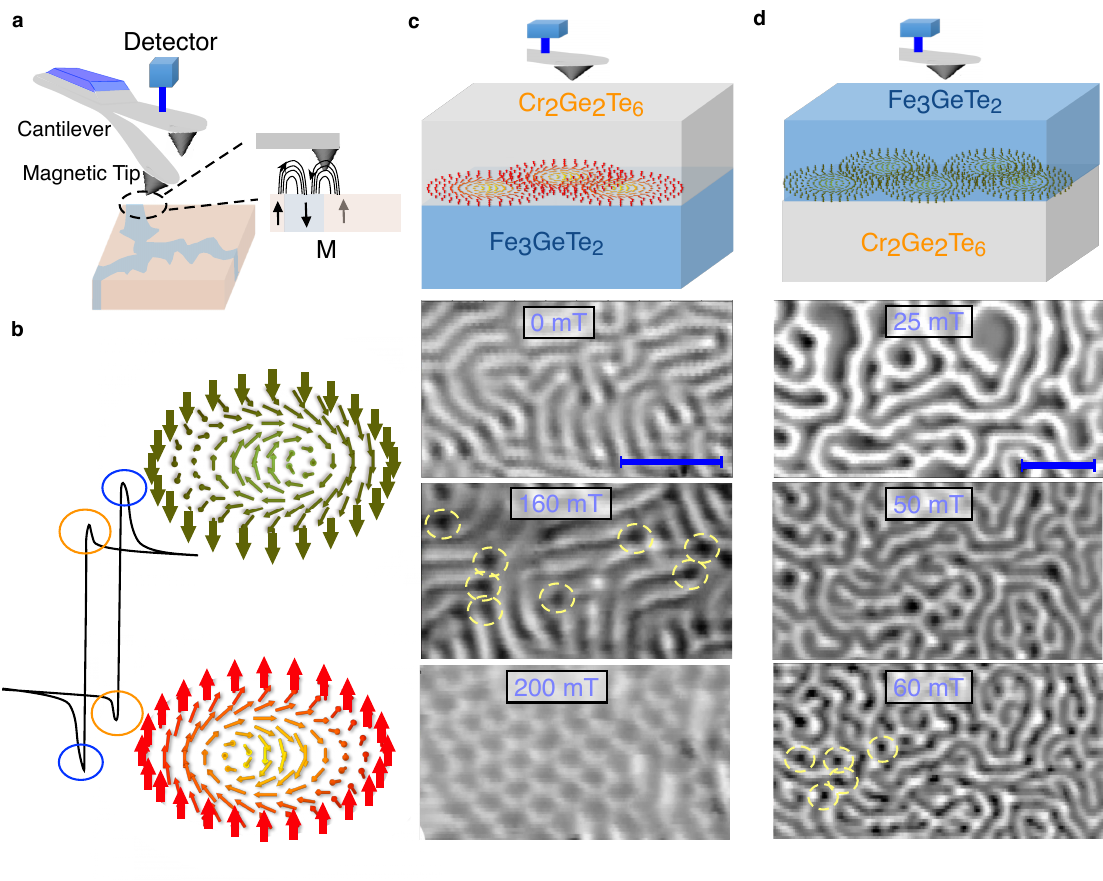}
\caption{\textbf{Skyrmions at both sides of Fe$_3$GeTe$_2$/Cr$_2$Ge$_2$Te$_6$ interface imaged by MFM}. \textbf{a}, Schematic illustration of MFM measurements. \textbf{b}, Two sets of THE signals observed for temperatures lower than 60 K. The blue circles signifying THE on FGT side and orange circles signifying THE on CGT side. The peak and dip features of THE resistivity indicating the skyrmions with opposite polarity. \textbf{c}, Skyrmion lattice observed on CGT side with a magnetic field of 200 \,mT and a temperature of 20 K.  Scale bar: 1 $\mu$m. \textbf{d}, Skyrmions on FGT side start to appear with a magnetic field of 50 mT at a temperature of 100 K. Scale bar: 1 $\mu$m. \label{Fig:Skyrmion}}
\end{figure}

To further examine the origin of THE arising at the FGT/CGT interface, we performed low-temperature MFM measurements under applied magnetic fields. In MFM, the magnetic stray field originating from nanoscale features on the surface of a sample is detected by a scanning nanoscale magnetic tip (Fig. \ref{Fig:Skyrmion}a). Note that in our topological Hall resistivity, two sets of kinks were observed, attributable to magnetic skyrmions on FGT side and CGT side, respectively (Fig. \ref{Fig:Skyrmion}b). In our MFM measurement geometry, the MFM signal characterizes the second order derivative of the out-of-plane component of the stray magnetic field, i.e., d$^2H_z/dz^2$. For samples with out-of-plane magnetization, MFM responds both inside individual domains and at the domain walls. Fig. \ref{Fig:Skyrmion}c shows a series of MFM images taken on CGT/FGT device (same device as in Fig. \ref{Fig:THE}b) with applied perpendicular magnetic fields at 20 K. As the field increases from zero, the image shows a labyrinth domain feature. Upon increasing the field to 160 mT, skyrmionic features started to form as indicated by the yellow circles in Fig. \ref{Fig:Skyrmion}c. At 200 mT, skyrmion lattice with a size of $\sim$130 nm was directly imaged. This skyrmion lattice disappears and the sample enters into a uniform single domain as the field is increased to 250 mT (see Supplementary Information Section B for skyrmion lattice at larger magnetic fields). This observed magnetic skyrmions are expected to occur at the interface close to CGT side for three reasons: (1) with CGT on top of FGT, MFM signals would be dominated by the magnetic fields generated by the CGT layer, (2) the magnetic signal from 5L FGT under multilayer CGT is too weak or screened to be observed, and (3) the formation of magnetic skyrmion lattices exactly matches the magnetic field which lead to the development of THE on CGT side (the dip at a positive magnetic field in Fig. \ref{Fig:THE}b and \ref{Fig:Skyrmion}b).  

With a reversed sample structure, i.e., FGT on top of CGT, MFM measurements show domain images in FGT side in Fig. \ref{Fig:Skyrmion}d. At 100 K, FGT shows similar labyrinth domain feature at 25 mT but with a larger domain width, compared to CGT case at 20 K. When the field was increasing, skyrmions started to appear, as indicated by the yellow circles. The skyrmion size was estimated to be $\sim$ 180 nm (100\,K and 60\,mT) and decreasing as the temperature went down. Magnetic skyrmions lattice at a lower temperature can be found in Supplementary Information Section C.

\begin{figure}
\centering{}
\includegraphics[width=0.6\columnwidth]{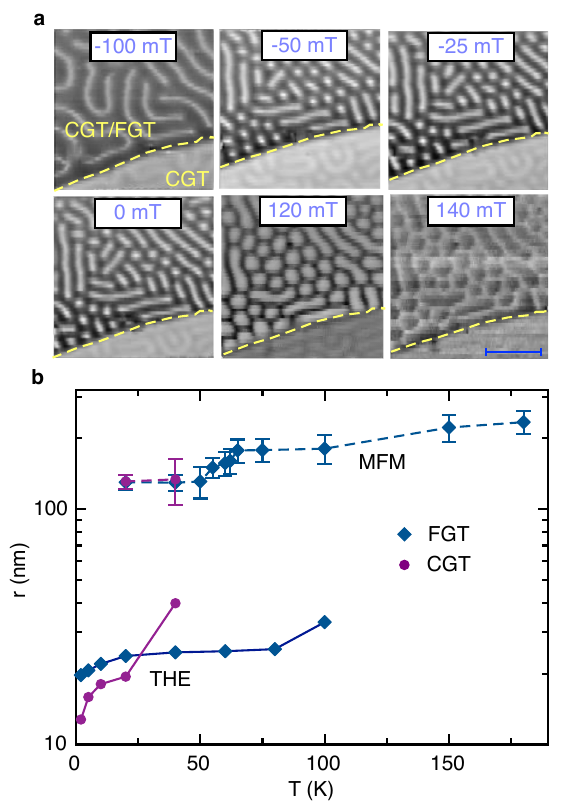}
\caption{\textbf{Magnetic skyrmions showing up as a function of magnetic field (even at zero field) and skyrmion size dependence on the temperature}. \textbf{a}, Skyrmion lattice on CGT side under various applied magnetic fields at 20 K after subjecting to -2 T. Scale bar: 1 $\mu$m. \textbf{b}, Skyrmion size extracted from topological Hall effect (an order-of-magnitude estimate) and from magnetic force microscopy images. The date points with dashed lines are from magnetic force microscopy measurements and the data points with solid lines are from topological Hall effects. \label{Fig:size}}
\end{figure}

The skyrmion lattice in CGT/FGT develops with the magnetic field. When sweeping from a negative field (-2 T) which is large enough to saturate CGT magnetization at 20 K, the skyrmions develop at $\sim$ -50 mT (Fig. \ref{Fig:size}a). Interestingly, the magnetic skyrmions are seen to persist when the applied magnetic field is turned off, in the remanent state. Emergence of magnetic skyrmion at 0 T magnetic field has been predicted to be a possible signature of Moir\'e skyrmions\cite{tong2018skyrmions}. The real reason and mechanism of zero-field magnetic skyrmions is not clear in our case at present. One should also consider the role of the exchange coupling between CGT and FGT, where FGT provides an effective field at the interface. When the magnetic field was further increased to 120 mT, skyrmion lattice emerged forming a hexagonal pattern. Further increasing the magnetic field to 140 mT, the skyrmion images switched to opposite contrast due to the flip of tip magnetization by the external magnetic field.

The skyrmion size was extracted from THE and the skyrmion images by MFM measurement (Fig. \ref{Fig:size}b). The skyrmion size for both FGT and CGT side decreases with decreasing temperature. The error bars for data points in the MFM measurement (dots and squares with dashed lines) are due to the variation of skyrmion size under applied magnetic field. The skyrmion size extracted from the THE (dots and squares with solid lines) offers an order of magnitude estimate (details of the size extraction can be found in Supplementary Information Section D), which is around one order of magnitude smaller than that from MFM measurement.

For an interface formed between CGT and FGT, we can also estimate the DMI energy from the strip domain width. The DMI in FGT is estimated to be, $|D|=0.31 \pm 0.13$ mJ m$^{-2}$ (details can be found in Supplementary Information Section E), a value similar to that reported in WTe$_2$/FGT system\cite{wu2020large}, a vdW heterostructure hosting skyrmion lattice (difference for the skyrmions in the heterostructure and skyrmion bubbles in the pristine centrosymmetric magnets can be found in Supplementary Information Section F).    

\subsection*{Summary and outlook}

We explored the ultrathin FGT/CGT heterostructures as a platform for hosting small and tunable two groups of magnetic skyrmions. In this system, the inversion symmetry breaking near the FGT/CGT interface gives rise to an emergent DMI, thereby creating robust magnetic skyrmions. By harnessing the interface between two vdW ferromagnets, we achieved magnetic skyrmions on two sides of an interface. 

The multi groups of skyrmions in FGT/CGT heterostructures allow versatility when designing and fabricating skyrmion-based devices. Due to the atomically thin nature of vdW materials and their easy control by electrical gating, one can foresee manipulating the skyrmions on each side with dual electric gating by which the skyrmion phase can be turned on or off. Furthermore, the twist angle between FGT and CGT may be tuned, where a new functionality could be added for manipulating skyrmions in atomically thin vdW heterostructures. With an accurate determination of twist angle between FGT and CGT layers, magnetic skyrmions from DMI or Moir\'e pattern may be differentiated. This versatile advantage offers a fertile playground for exploring emergent phenomena arising from hybrid interfaces, leading to magnetic skyrmions with additional functionalities.

\subsection*{Methods}

\subsubsection*{Growth of Fe$_3$GeTe$_2$ and Cr$_2$Ge$_2$Te$_6$}

High quality single crystals of FGT were grown by a typical chemical vapor transport (CVT) method.  The stoichiometric amounts of high purity elements (99.999\% Fe, 99.999\% Ge and 99.999\% Te from Alfa Aesar) along with 2 mg cm$^{-3}$ iodines the transport agent were placed in a quartz ampoule and sealed under vacuum.  The ampoule was further placed in a horizontal 2-zone furnace over a temperature gradient from 750$^{\circ}$C to 700$^{\circ}$C and kept at that condition for 2 weeks. The single crystals with a tabular shape in the $ab$-plane grew in the low-temperature zone.

High quality CGT single crystals were synthesized from high purity elements (99.99\% Cr, 99.99\% Ge and > 99.99\% Te from Alfa Aesar). A mixture of materials with an optimized ratio of Cr : Ge : Te equivalent to 10 : 13.5 : 76.5, was sealed in evacuated quartz ampoules and heated at 1000$^{\circ}$C for one day, followed by slow cooling to 450$^{\circ}$C for a period of 90 hours.   

\subsubsection*{MFM measurement}

The MFM measurements were performed in a home-built low-temperature scanning probe microscope using commercial MFM probes (Bruker MESPV2) with a spring constant of 3 N m$^{-1}$, a resonance frequency at $\sim$75 kHz, and a Co-Cr magnetic coating. MFM images were taken in a constant height mode with the tip scanning plane at $\sim$100 nm above the sample surface. The MFM signal, the change in the resonance frequency, is measured by a Nanonis SPM Controller using a phase-lock loop. The magnetic moment of the probe is nominally in the order of $10^{-16}$ A\,m$^{2}$.

\subsubsection*{Magnetotransport measurement }

Hall-bar bottom electrodes with dimension of $3.5\,\mu\textrm{m}\times0.5\,\mu\textrm{m}$ were fabricated with e-beam lithography for the transport measurements. Systematically altering experimental variables such as temperature and magnetic field, in addition to multiple lock-in amplifiers and sourcemeter enabled comprehensive and high-sensitivity transport measurements in all the devices, using a commercial PPMS from Quantum Design (2 K, $\pm$ 9T). 

\subsection*{Data availability Statement}

The data that support the findings of this study are available from the corresponding authors upon request. 

\subsection*{References  }
\bibliographystyle{naturemag}
\bibliography{C-FGT_v18.bib}
\break

\subsection*{Acknowledgements:}

The authors are grateful for discussion with Gen Yin from Georgetown University. The transport measurements in this work are supported by the Army Research Office (ARO) program under contract W911NF-15-1-10561, National Science Foundation (NSF) with Award Nos. 1935362 and 1909416, CIQM-NSF DMR-1231319, and ARO grant W911NF-20-2-0061. H. C. was sponsored by the Army Research Laboratory under cooperative agreement no. W911NF-19-2-0015.
We are also grateful to the support from the NSF (DMR-1411085) and Department of Energy, Office of Science under Award No. DE-SC0020221. MFM measurements are supported by NSF DMR-2145735.

\subsection*{Author Contributions}

Y.W. conceived and designed the experiment. K.L.W. and J.S.M. supervised the project. W.W. grew CGT and FGT bulk crystals. Y.W. performed the fabrication and carried out the transport measurements. B.F. and Y.C. implemented the MFM measurements. Y.Z., C.W. and X.H. prepared the bottom metal electrodes. H.C., Y.H. and A.L. helped with experimental setup. Y.W. and J.S.M. wrote the manuscript with contributions from all the authors. 

\subsection*{Competing interests}

The authors declare no competing interests.\newpage{}

\end{document}